\documentstyle[preprint,aps,12pt]{revtex}
\begin{document}
\preprint{YITP/K-1116}
\title{Spin-$S$ generalization of fractional exclusion statistics}
\author{ Takahiro Fukui \cite{JSPS}}
\address{Yukawa Institute for Theoretical Physics, Kyoto University,
Kyoto 606-01, Japan}
\author{Norio Kawakami}
\address{Department of Material and Life Science,
and Department of Applied Physics,\\
Osaka University, Suita, Osaka 565, Japan}
\author{Sung-Kil Yang}
\address{Institute of Physics, University of Tsukuba, Ibaraki 305, Japan}
\date{July, 1995}
\maketitle
\begin{abstract}
We study fractional exclusion
statistics for quantum systems with SU(2) symmetry
(arbitrary spin $S$), by generalizing the thermodynamic
equations with squeezed strings proposed by Ha and Haldane.
The bare hole distributions as well as
the statistical interaction defined by an
infinite-dimensional matrix specify
the universality class.
It is shown that the system is described by the level-$2S$
WZW model and has a close relationship to non-abelian
fractional quantum Hall states.
As a low-energy effective theory, the sector of
{\it massless} Z$_{2S}$ parafermions
is extracted, whose statistical interaction
is given by a finite-dimensional matrix.
\end{abstract}
\pacs{75.10.Jm, 73.40.Hm}

Fractional exclusion statistics proposed by Haldane\cite{HAL}
has attracted much current interest.
In particular, the notion of {\it ideal} exclusion statistics
is useful to characterize elementary excitations,
which naturally interpolates free fermions and free
bosons\cite{HAL,MURTHY,WU,BERNARD}.
The ideal exclusion statistics is featured by
statistical interaction $g_{\mu\nu}$
defined as,
\begin{equation}
\frac{\partial D_\mu}{\partial N_\nu}=-g_{\mu\nu},
\label{SI}
\end{equation}
where $g_{\mu\nu}$ depends on the species $\mu$ and  $\nu$,
but not on the momentum $k_j$.
Here $D_\mu$ and $N_\mu$ denote, respectively,
the number of holes and particles with species $\mu$.
A typical system in which ideal exclusion statistics
is realized is quantum $1/r^2$
models in one dimension\cite{ISM,ISMD}.

The ideal exclusion statistics
has a close relationship\cite{FUKUI,FUKUITWO} to
{\it abelian} Chern-Simons theory
for the fractional quantum Hall effect (FQHE)\cite{WENZEE}.
It turns out that the
statistical interaction matrix plays the same role as the
topological-order matrix in FQHE, and hence the
multicomponent model in ref.
\cite{FUKUI} may serve as a model for the edge states of the FQHE,
namely, its low-energy behavior
is described by multiple $c=1$ conformal field theories (CFTs).
In this sense, we may call such statistics {\it abelian fractional
exclusion statistics}.
In theories of the FQHE, another interesting class of
FQH states has been proposed, in which quasiparticles obey
{\it non-abelian} statistics\cite{MOORE,BLOCKWEN,WENONE,WENTHR,WENTWO}.
The corresponding theory is given
by, for example, non-abelian Chern-Simons theory, and is
closely related to $c>1$ CFT. From these observations,
a natural question arises, i.e.
how can one generalize fractional
exclusion statistics to non-abelian cases?

We address this question in this paper, by applying
the idea of exclusion statistics to SU(2) higher-spin models in
one dimension. We propose basic equations
for ideal statistics with arbitrary spin based on the idea of squeezed
strings introduced in the $S=1/2$ model \cite{HA}.
We then study the thermodynamics and show that the
critical behavior is described by the level-$2S$ SU(2) WZW model.
It is found that
{\it in order to formulate non-abelian statistics, the bare
hole distribution also plays a vital role in addition to the statistical
interaction}.
We next deduce explicitly
the statistical interaction of massless
Z$_{2S}$ parafermion which characterizes the nontrivial properties of
non-abelian statistics.
The relation to non-abelian FQHE is mentioned briefly.

To begin with, let us briefly review the ideal spinon description
of the quantum system with $S=1/2$\cite{HAL}.
In this case, ideal statistics is modeled by
lattice bosons with  statistical interaction $g=2$,
reflecting level-1 SU(2) Kac-Moody symmetry.
Recall that the ideal
statistical interaction $g_{\mu\nu}$ is
simply related to the two-body phase shift $\phi$ of the scattering
matrix in one dimension through
$\phi(k_i-k_j)=\pi \, (g_{\mu\nu}-\delta_{\mu\nu}) {\rm sgn}(k_i-k_j)$
\cite{HAL,WU,BERNARD}.
Note that the phase shift with step function is inherent in
ideal statistics.
Thus the two-body phase shift in the $S=1/2$ model reads
$\phi=\pi \, {\rm sgn} (k_i-k_j)$. Using this phase shift one could naively
write down the Bethe equation to determine the spectrum for the periodic
chain. The resulting equation, however, does not yield the right number of
eigenstates. To overcome this difficulty Ha and Haldane introduced
the fictitious string solutions called ``squeezed strings''
$k_j^{\mu} \in {\bf R}$ ($\mu=1,2,\cdots$) \cite{HA}.
The Bethe equation now becomes
the coupled equations for strings.
What is remarkable is that these string solutions give
not only complete degeneracy of states
but also the correct eigenvalues
even for finite systems.
This method thus provides precise thermodynamics of the
free spinon gas obeying ideal exclusion statistics.

We now wish to generalize this description to the higher spin model.
We expect that the basic equation
for ideal statistics may have
a similar structure to the Bethe equation for integrable models.
Examining carefully the Bethe equation for
solvable Heisenberg chain with higher spin\cite{TAKHTAJIAN,BABUJIAN},
we are led to propose
the following basic equation, which
generalizes the Ha-Haldane equation to arbitrary spin-$S$,
\begin{equation}
d_\mu^{(S)} Nk_i^\mu =2\pi I_i^\mu +\pi\sum_{\nu ,j}c_{\mu\nu}
{\rm sgn} (k_i^\mu -k_j^\nu )
\label{HSBAE}
\end{equation}
for $\mu=1,2, \cdots $, where $d_\mu^{(S)} =\min (\mu ,2S)$,
$c_{\mu\nu}=G_{\mu\nu}-\delta_{\mu\nu}$ with
\begin{equation}
G_{\mu\nu}=2\min (\mu ,\nu ) ,
\label{G}
\end{equation}
and $N$ is the number of sites.
Each $k$ is defined in the region $k\in [-\pi ,\pi ]$.
In the above equation, we have introduced the squeezed
string solutions $k_j^{\mu}$,
which are coupled with each other by the matrix
$c_{\mu\nu}$.
The statistical interaction of this model
is given by the infinite dimensional matrix $G$.
The total energy is given by
$E=\sum_{\mu ,i}\epsilon_\mu^0(k_i^\mu )$, where
\begin{equation}
\epsilon_\mu^0(k)=d_\mu^{(S)}\epsilon^0(k), \quad
\epsilon^0(k)=\frac{1}{2}(k^2-\pi^2).
\label{DIS}
\end{equation}
Here we have assumed  the quadratic dispersion of the
energy, which is indeed the case for integrable  $1/r^2$
models\cite{ISMD}, though the following analysis is not
sensitive to the form of $\epsilon^0$.
Note that the r.h.s. of eq.(\ref{HSBAE}) takes the same form as
the Ha-Haldane equation.
Thus the effect of higher spin is incorporated through
the factor $d^{(S)}$ in eqs.(\ref{HSBAE}) and (\ref{DIS}),
whose role is crucial to identify the universality class
of the  system. We will show momentarily that
$d^{(S)}$ is related to the bare hole distribution.

In the thermodynamic limit, eq.(\ref{HSBAE}) leads to
the coupled equations for the distribution functions of particles
($\rho$) and holes ($\rho^{(h)}$),
\begin{equation}
\rho_\mu(k) +\rho_\mu^{(h)} (k)=\frac{1}{2\pi}d_\mu^{(S)}-
\sum_\nu c_{\mu\nu}\rho_\nu (k) .
\label{DISTR}
\end{equation}
The equilibrium distributions are then determined
by minimizing the free energy $\delta F=0$, which gives
\begin{equation}
\epsilon_\mu (k) =\epsilon_\mu^0 (k) +\mu H+
T\sum_\nu c_{\mu\nu}\ln (1+e^{-\epsilon_\nu (k) /T}) ,
\label{DE}
\end{equation}
where $H$ is a magnetic field, and the dressed
energy $\epsilon$ is introduced  by
$\rho^{(h)}/\rho =e^{\epsilon /T}$. The free energy per site
$f=F/N$ at equilibrium is then given by
\begin{equation}
f=-HS-\frac{T}{2\pi}
\sum_\mu d_\mu^{(S)}\int_{-\pi}^\pi dk\ln (1+e^{-\epsilon_\mu /T}),
\end{equation}
which  can be rewritten by using eq.(\ref{DE}),
\begin{equation}
f=-\frac{\pi^2}{3}S-
\frac{T}{4\pi}\int_{-\pi}^\pi dk\ln (1+e^{\epsilon_{2S}/T}) .
\label{FE}
\end{equation}
Eqs.(\ref{DISTR}), (\ref{DE}) and (\ref{FE}) are the basic equations
for describing thermodynamics of the systems
obeying non-abelian exclusion statistics with
an arbitrary  spin $S$.

Let us now examine low-temperature properties. First, by using
$G_{\mu\nu}^{-1}=
(2\delta_{\mu\nu}-\delta_{\mu ,\nu +1}-\delta_{\mu +1,\nu})/2$,
eq.(\ref{DE}) can be converted into
\begin{equation}
\frac{\epsilon_\mu}{T}=\frac{1}{2}(L_{\mu -1}+L_{\mu +1})
+\frac{\epsilon^0}{2T}\delta_{\mu ,2S} ,
\label{DEE}
\end{equation}
where $L_\mu =\ln (1+e^{\epsilon_\mu /T})$.
We can see from this equation that $\epsilon_\mu\ge 0$, while
$\epsilon_{2S}$ can change its sign at some value $k=\pm Q$
(pseudo-Fermi points).
At $T=0$, eq.(\ref{DE}) reduces to
\begin{eqnarray}
\epsilon_\mu (k)
&=&\epsilon_\mu^0 (k) +\mu H-G_{\mu 2S}\epsilon_{2S}^{(-)} (k)
\qquad {\rm for}\quad \mu\ne 2S,\label{DEZEROD}\\
\epsilon_{2S}^{(+)} (k)
&=&\epsilon_{2S}^0 (k) +2SH-G_{2S 2S}\epsilon_{2S}^{(-)} (k) ,
\label{DEZERO}
\end{eqnarray}
where $\epsilon_{2S}^{(+)}=\theta (\epsilon_{2S}(k))$
and $\epsilon_{2S}^{(-)}=-\theta (-\epsilon_{2S}(k))$,
with  $\theta(x)=1\, (0)$ for $x>0\, (x<0)$.

We start with  the case of $H=0$ to simplify the
discussions. At $T=0$, we immediately find
\begin{equation}
\epsilon_\mu (k) =\frac{\epsilon^0}{2}\delta_{\mu ,2S},
\end{equation}
which implies that the ground state consists of
the sea of $2S$-strings with the pseudo-Fermi points
$\pm Q$ with $Q=\pi$. At low temperatures, we can thus
see from (\ref{FE}) that the most
important contributions to
the free energy come from the vicinity of $k=\pm Q$.
For example, in the region close to the right Fermi point,
$[\pi -\epsilon ,\pi ]$,
the driving term in eq.(\ref{DEE}) is
approximated by $\pi (k-\pi )/2T\equiv -e^{-\theta}$ where
$\theta$ is defined in the region
$\theta\in [-\ln (\pi\epsilon /2T),\infty )$.
In the limit $T \rightarrow 0$, we
can set $\theta\in (-\infty ,\infty )$. We get from eq.(\ref{DEE})
\begin{equation}
\varphi_\mu =\frac{1}{2}(L_{\mu -1}+L_{\mu +1})
-e^{-\theta}\delta_{\mu ,2S},
\label{CONVDE}
\end{equation}
where $\varphi_\mu (\theta )=\epsilon_\mu /T$.
The free energy is then given by
\begin{eqnarray}
f&=&-\frac{T^2}{2\pi^2}\int_{-\infty}^\infty d\theta
e^{-\theta}L_{2S}\nonumber\\
&=&-\frac{T^2}{4\pi^2}\sum_{\mu =1}^\infty\int_{-\infty}^{\infty}d\theta
(\varphi_\mu 'L_\mu -\varphi_\mu L'_\mu ) ,
\end{eqnarray}
where we dropped the contribution from the ground-state energy.
In order to evaluate this integral, we must solve
eq.(\ref{CONVDE}) in the asymptotic
region $\theta\rightarrow\pm\infty$.
We find\cite{BABUJIAN}
\begin{eqnarray}
\omega_{\mu -}&=&\left\{
\begin{array}{ll}\left(\displaystyle{
\frac{\sin\displaystyle{\frac{(\mu +1)\pi}{2S+2}}}
{\sin\displaystyle{\frac{\pi}{2S+2}}}}
\right)^2 &\quad{\rm for}\quad 1\le \mu\le 2S-1 ,\\
(\mu -2S+1)^2 &\quad{\rm for}\quad 2S\le\mu ,
\end{array}\right.\\
\omega_{\mu +}&=&(\mu +1)^2 ,
\end{eqnarray}
where $\omega_{\mu\pm}=e^{\varphi_\mu (\pm\infty )}+1$.
Consequently  we have the low-energy expansion of
the free energy,
\begin{equation}
f=-2\frac{T^2}{2\pi^2}\sum_{\mu =1}^{2S}{\cal L}(\omega_{\mu -}^{-1})
=-\frac{\pi T^2}{6v}\frac{3k}{k+2},
\end{equation}
where the factor 2 in the first expression
is due to the contributions from both the left and right Fermi points,
${\cal L}(x)$ is Rogers dilogarithm function
defined by
\begin{equation}
{\cal L}(x)=-\frac{1}{2}\int_{0}^{x}dt
\left[\frac{\ln (1-t)}{t}+\frac{\ln t}{1-t}\right] ,
\end{equation}
$k=2S$ and $v=\pi$ is the Fermi velocity of the $2S$-string sector.
Here we have used dilogarithm sum rules\cite{KIRILLOV}.
According to the finite-size scaling law\cite{BLOTE,AFFLECK},
we see that the critical behavior of
the present system can be described by the SU(2)$_{2S}$ WZW model.
Namely, statistical interaction $G$ in eq.(\ref{G}) with additional
$d^{(S)}$ in eqs.({\ref{HSBAE}) and (\ref{DIS})
results in  the central charge,
\begin{equation}
c=\frac{3k}{k+2},
\label{C}
\end{equation}
with $k=2S$. This result implies that the
universality class of the present system is characterized
not only by the statistical interaction $G$ but also by the
factor $d^{(S)}$.
Hence, $d^{(S)}$ also plays a role in
formulating ideal statistics for general $S$. Let us check
how $d^{(S)}$ enters in the definition
of exclusion statistics (\ref{SI}).
To see this, we integrate (\ref{SI}) to obtain\cite{BERNARD}
\begin{equation}
D_\mu =-\sum_{\nu}g_{\mu\nu}N_\nu +D^0_\mu.
\end{equation}
One can see that  $d^{(S)}$ is
directly related to the integration
constant $D^0_\mu$ such that $D_\mu^0=Nd_\mu^{(S)}/2\pi$.
This term can be interpreted as the bare distribution of holes when
there are no particles in the system.
Therefore,
{\it the bare hole distributions as well as
the statistical interaction
are the fundamental elements which
characterize fractional exclusion statistics
for non-abelian cases}.
This point has not been seriously taken into
account so far, because
the above problem does not happen for abelian cases.

We now derive low-energy effective theories.
In the low-energy region, we can
explicitly separate the theory into the abelian $c=1$ sector and
the $c=2(k-1)/(k+2)$ Z$_k$ sector.
This analysis enables us to  naturally formulate
ideal statistics with Z$_{2S}$ symmetry.
We will extract the Z$_{2S}$ parafermion sector following
Tsvelik\cite{TSVELICK}.
First, we switch on magnetic fields to
deduce effective equations for modes $1\le\mu\le 2S-1$.
At $T=0$, we have from eq.(\ref{DEZEROD}),
\begin{equation}
\epsilon_\mu (k) =\left\{\begin{array}{ll}
\displaystyle{\frac{\mu}{2S}}\epsilon_{2S}^{(+)}
&\quad{\rm for}\quad \mu < 2S,\\
(\mu -2S)H+\epsilon_{2S}^{(+)}& \quad{\rm for}\quad 2S<\mu,\end{array}\right.
\label{PARADIS}
\end{equation}
and from eq.(\ref{DEZERO}),
\begin{eqnarray}
\epsilon_{2S}^{(+)} (k) &=&\left\{
\begin{array}{ll}2S(\epsilon^0(k)+H)
&\qquad{\rm for}\quad Q<|k|\le\pi ,\\
0&\qquad{\rm for}\quad |k|<Q,\end{array}\right.\\
\epsilon_{2S}^{(-)} (k) &=&\left\{
\begin{array}{ll}
0&\qquad{\rm for}\quad Q<|k|\le\pi ,\\
\displaystyle{\frac{1}{2}}
(\epsilon^0(k) +H)&\qquad{\rm for}\quad |k|<Q, \end{array}\right.
\end{eqnarray}
where $\epsilon_{2S}(\pm Q)=0$ with $Q=\sqrt{\pi^2-2H}$. From these equations
we see that the $\mu < 2S$ modes (parafermions) as well as $\mu =2S$
mode (U(1)) are still  {\it massless},
while those for $2S<\mu$ get massive.
This is a novel feature which has not been observed
for the integrable spin chain\cite{TAKHTAJIAN,BABUJIAN}
in which all modes become massive under magnetic fields except
for $2S$-string sector\cite{TSVELICK}. The existence of
massless parafermions is closely related to
ideal statistics, i.e. it directly reflects
the step-function of the phase shift.
If the deviation from ideal  statistics is introduced,
the parafermion sector immediately
becomes massive, as will be discussed later.
In this sense, the present model is regarded
as an ideal model for the {\it massless spin chain}
with arbitrary $S$.

Let us proceed to calculate the distribution functions.
Subtracting eq.(\ref{DISTR})
for $\mu <2S$ from that for $\mu =2S$, we obtain
\begin{equation}
\rho_\mu^{(h)}+\rho_\mu =\frac{\mu}{2S}\rho_{2S}^{(h)}-
\sum_{\nu =1}^{2S-1}(\widetilde G_{\mu\nu}-\delta_{\mu\nu})\rho_\nu
\label{PARABE}
\end{equation}
with $\mu ,\nu =1,...,2S-1$, where
\begin{equation}
\widetilde G_{\mu\nu}=G_{\mu\nu}-\frac{\mu\nu}{S}.
\label{GTIL}
\end{equation}
The modes $\mu <2S$ still couple to the $2S$-string sector.
Since the U(1) and Z$_{2S}$ sectors are both massless, it seems
not easy  to separate them. Nevertheless, if
we replace $\rho_{2S}^{(h)}$ in the above
equation by the ground-state distribution,
\begin{equation}
\rho_{2S}^{(h)}=\left\{\begin{array}{ll}
\displaystyle{S/\pi}&\quad{\rm for}\quad Q<|k|\le\pi ,\\
0&\quad{\rm for}\quad |k|<Q ,\end{array}\right.
\label{gsdistri}
\end{equation}
we then get desirable equations describing  Z$_{2S}$ parafermions from
eq.(\ref{PARABE}) with the dispersion (\ref{PARADIS}).
This can be easily confirmed by examining the thermodynamics again
based on these equations and computing the
central charge, which in fact results in $c=2(k-1)/(k+2)$.
Hence we conclude that (\ref{PARABE}) with (\ref{gsdistri})
describes  ideal exclusion statistics for
Z$_{2S}$ parafermions.  Their statistical interaction is
given by $(2S-1)\times (2S-1)$ matrix $\widetilde G$ in
eq.(\ref{GTIL}).
Note that $\widetilde G$ is the inverse of
the A$_{2S}$ Cartan matrix $C$:
\begin{equation}
C\widetilde G=2 I.
\end{equation}
Therefore it can be expressed by the incidence matrix $l$;
$\widetilde G-I=l(2-l)^{-1}$.
Having this expression here is natural since it
has already appeared in the thermodynamic
Bethe ansatz  for the massive perturbation theory of CFT whose
ultra-violet behavior is described by Z$_{N}$
parafermions\cite{KLASSEN}.

It is instructive to consider a question why the present system
has massless parafermions even in magnetic  fields
in contrast to the ordinary solvable spin chain models
\cite{TAKHTAJIAN,BABUJIAN,TSVELICK}.
To clarify this point, let us take equations,
\begin{equation}
d_\mu^{(S)}Nk(\lambda_i^\mu )=2\pi I_i^\mu +\sum_{\nu ,j}
\Xi_{\mu\nu} (\lambda_i^\mu -\lambda_j^\nu ),
\end{equation}
where $k(\lambda )=2\arctan\lambda$, $\Xi_{\mu\nu}
=\theta_{|\mu -\nu |}+2\theta_{|\mu -\nu |+2}+...+
2\theta_{\mu +\nu-2}+\theta_{\mu +\nu}$ (for $\mu =\nu$ the first term
is omitted) with $\theta_\mu (\lambda )=2\arctan (2\lambda /
\Gamma\mu )$. This equation recovers
(\ref{HSBAE}) in the $\Gamma\rightarrow 0$ limit. Also, one notices
that r.h.s. is the same as the Bethe equation for
the solvable XXX spin chain for $\Gamma=1$.
Using this equation, we can derive dressed energies for $\mu <2S$ modes,
\begin{equation}
\epsilon_\mu (\lambda )=\int_{-\infty}^\infty d\lambda 'K_\mu
(\lambda -\lambda ')\epsilon_{2S}^{(+)}(\lambda '),
\end{equation}
where
\begin{equation}
K_\mu (\lambda )=\frac{1}{N\Gamma}\frac{\sin(\pi\mu/N)}
{\cosh(2\pi\lambda/N\Gamma)+
\cos(\pi\mu/N)},
\end{equation}
apart from a term which disappears with $\Gamma\rightarrow 0$.
For $\Gamma=1$, this expression coincides with that for
the XXX spin chain derived by Tsvelik, giving  a massive dispersion
$\epsilon_\mu (\lambda )\propto\sin (\pi\mu /N)\cosh (2\pi\lambda /N)$
for small $\lambda$\cite{TSVELICK}.
Contrary to this, in the $\Gamma\rightarrow 0$
limit, $K_\mu (\lambda )\rightarrow
(\mu /N)\delta (\lambda )$, and this gives
massless dispersions (\ref{PARADIS}).
One can thus see that massless parafermions are realized only
for ideal statistics, and  infinitesimal deviation from
it may bring about massive parafermions
in magnetic fields.

Finally, let us briefly mention
the relation to the edge states of the
FQHE. As proposed by Moore and Read\cite{MOORE}, and Block and Wen
\cite{BLOCKWEN,WENONE,WENTHR,WENTWO},
there are several models for non-abelian FQH states. An example is
the FQH state with U(1)$\times$ SU($m$)$_n$
symmetry\cite{WENONE,WENTWO}.
The present results can describe the behavior
of the non-abelian SU(2)$_n$ sector of the above edge states.
Another interesting example is the FQH state with U(1)$\times$Z$_N$
model proposed by Moore and Read\cite{MOORE,BLOCKWEN,WENTHR,WENTWO}}.
The ``Pfaffian" part of Z$_N$ is essential for this non-abelian FQH state.
Our result for Z$_{2S}$ symmetry extracted from the SU(2)$_{2S}$
case may describe the above sector
in the edge state.  It may be interesting to generalize
the present discussions to describe non-abelian
statistics with more complicated symmetries.

This work is supported by the Grant-in-Aid from the Ministry of
Education, Science and Culture, Japan.

\end{document}